\documentstyle[aps,prb,epsfig,floats]{revtex}

\newcommand{\del}{\partial}
\newcommand{\pas}{{{\bf p} \alpha \sigma}}
\newcommand{\kas}{{{\bf k} \alpha \sigma}}
\newcommand{\ppbs}{{{\bf p}' \beta \sigma}}
\newcommand{\kgs}{{{\bf k} \gamma \sigma}}
\newcommand{\kpds}{{{\bf k}' \delta \sigma}}
\newcommand{\kgsb}{{{\bf k} \gamma \bar{\sigma}}}
\newcommand{\kpdsb}{{{\bf k}' \delta \bar{\sigma}}}

\newcommand{\eps}{{\varepsilon}}
\newcommand{\eas}{{E \alpha \sigma}}
\newcommand{\epbs}{{E' \beta \sigma}}
\newcommand{\epsgs}{{\eps \gamma \sigma}}
\newcommand{\epsgsb}{{\eps  \gamma \bar{\sigma}}}
\newcommand{\epbsb}{{E' \beta \bar{\sigma}}}
\newcommand{\epspds}{{\eps + \omega,\delta \sigma}} 
\newcommand{\epspdsb}{{\eps + \omega,\delta \bar{\sigma}}} 
\newcommand{\mps}{M^{\pas;\;\kgs}_{\ppbs;\;\kpds}}
\newcommand{\mks}{M^{\pas;\;\kgs}_{\kpds;\;\ppbs}}
\newcommand{\mpsb}{M^{\pas;\;\kgsb}_{\ppbs;\;\kpdsb}}

\newcommand{\mes}{M^{\eas;\;\epsgs}_{\epbs;\;\epspds}}
\newcommand{\mepss}{M^{\eas;\;\epsgs}_{\epspds;\;\epbs}}
\newcommand{\mesb}{M^{\eas;\;\epsgsb}_{\epbs;\;\epspdsb}}
\newcommand{\mepssb}{M^{\eas;\;\epsgsb}_{\epspds;\;\epbsb}}

\newcommand{\dels}{\delta(E_\pas + E_\kgs - E_\ppbs - E_\kpds)}
\newcommand{\delsbp}{\delta(E_\pas + E_\kgsb - E_\ppbs - E_\kpdsb)}

\newcommand{\tph}{\frac{2\pi}{\hbar}}
\newcommand{\half}{\frac{1}{2}}

\begin{document}
\title{Dynamics of Excited Electrons in Copper and Ferromagnetic
  Transition Metals: Theory and Experiment} 
\author{R.~Knorren$^*$ and K.H.~Bennemann} 
\address{Institut f\"ur Theoretische Physik, Freie
  Universit\"at Berlin, Arnimallee 14, D-14195 Berlin, Germany}
\author{R.~Burgermeister}
\address{Laboratorium f\"ur Festk\"orperphysik, ETH Z\"urich, 8093
  Z\"urich, Switzerland}
\author{M.~Aeschlimann$^{\dag}$}  
\address{Laboratorium f\"ur Technische Chemie, ETH Z\"urich, 8092
  Z\"urich, Switzerland}
\date{\today} 
\maketitle
\begin{abstract}
Both theoretical and experimental results for the dynamics of
photoexcited electrons at surfaces of Cu and the ferromagnetic
transition metals Fe, Co, and Ni are presented. A model for the
dynamics of excited electrons is developed,  which is based on the
Boltzmann equation and  includes effects of photoexcitation,
electron-electron scattering, secondary electrons  (cascade and Auger
electrons),  and transport of excited carriers out of the detection
region.  From this we determine the time-resolved two-photon
photoemission (TR-2PPE). Thus a direct comparison of calculated
relaxation times with  experimental results by means of TR-2PPE
becomes possible. The comparison indicates that the magnitudes of the
spin-averaged relaxation time $\tau$ and of the ratio
$\tau_\uparrow/\tau_\downarrow$ of majority and minority relaxation
times for the  different ferromagnetic transition metals result not
only from density-of-states effects, but also from different Coulomb
matrix elements $M$. Taking $M_{\rm Fe}>M_{\rm Cu}>M_{\rm Ni}=M_{\rm
Co}$ we get reasonable agreement with experiments.
\end{abstract}

\section{Introduction}
The dynamics of excited electrons at metal surfaces has been  studied
intensively over the last few years. It is of high interest to
understand the dynamics of non-equilibrium 
electrons in different metals on a femtosecond timescale and its
influence on chemical reactions at surfaces, for example.
From the relaxation of hot electrons in ferromagnets one
may also learn about the decay of transient magnetization and about
spin-selective transport and tunneling.
 
Short laser pulses of 15--50 fs duration 
and the pump-probe technique have made possible the study
of electronic dynamics on ultrashort timescales comparable to typical
electron-electron interaction timescales, which in metals are of the
order of 5--50 fs for excitation energies of about 1--2 eV.

The aim of 2PPE experiments is to
study the relaxation of single excited electrons. 
A relaxation time is extracted from the width of the 2PPE
signal as a function of the delay time between pump and probe
pulses.\cite{Schmuttenmaer94} 
In the earlier
experiments,\cite{Schmuttenmaer94,Hertel96} the relaxation time was
interpreted as the lifetime of a single excited electron due to
the Coulomb interaction. It was compared to the theoretical result from
Fermi-liquid theory (FLT).\cite{Quinn62} The order of magnitude and the
energy dependence of the FLT lifetime 
were in good agreement with the experimental
results.\cite{Schmuttenmaer94,Hertel96,Aeschlimann96,Ogawa97a}
However, it was realized that additional physical effects such as 
the transport of excited
electrons out of the detection region and the secondary-electron
cascade come into play.\cite{Schmuttenmaer94,Aeschlimann96,Knoesel96} It
was noted that in 2PPE one  generally observes the relaxation of
a nascent photoexcited electron  population and not only the lifetime
of a single excited electron. For low excitation energies,
electronic lifetimes are 
longer than a few tens of fs, and ballistic transport 
leads to a removal of electrons from the probed region.\cite{Hohlfeld98} 
This is indistinguishable from a stronger
electronic decay. Furthermore, under certain
conditions the measured relaxation time shows a surprising non-monotonous
feature which depends on the photon energy of the exciting laser, which
cannot be explained by a single-electron lifetime and
transport.\cite{Pawlik97,Cao97,Knoesel98}
It was pointed out that the holes left behind in
the excitation can influence the 
observed relaxation time of hot electrons.\cite{Knoesel98,Sakaue99}
One explanation was that secondary electrons generated by the
filling of holes (Auger electrons) are responsible for the
non-monotonous behavior.\cite{Knoesel98}
However, the contribution of Auger electrons to the relaxation time
has raised some controversy. Petek {\it et al.} have argued that Auger
electrons do not make a significant contribution to the observed 2PPE
signal and to the relaxation time at high intermediate-state energies
$E-E_F>1.5-2\rm\ eV$.\cite{Petek97,Petek99}

The 3$d$ transition metals have not been as intensely studied as the
noble metals. However, they offer several interesting features which
make it worth to study them in detail. They offer the opportunity to
study 
spin-dependent interactions if the spin of the emitted electrons is
measured.\cite{Aeschlimann97,Aeschlimann98} Furthermore, the closeness
of the $d$ bands to the Fermi energy 
makes it possible to study electronic interactions not only for
free-electron-like states, but also for the more localized
$d$-electron states.

On the theoretical side, the effect of the density of states (DOS) on the
lifetime and the influence of secondary electrons in photoemission
from transition metals have been addressed by Penn {\it et
al.}\cite{Penn85} Using a similar approach, Zarate {\it et
al.}\cite{Zarate99} have calculated low-energy-electron lifetimes in
noble metals and ferromagnetic Co. First-principles lifetime calculations
have been performed
for image-potential states \cite{Chulkov98} and for bulk states in
alkali and noble metals.\cite{Campillo99,Schoene99} The lifetime is obtained
from the inverse of the imaginary part of the self-energy. As in
FLT, the lifetime calculated
in the above works is a single-electron lifetime. Due to the
additional effects of secondary electrons and transport in 2PPE
experiments, it is difficult to compare these theoretical results with
the relaxation times measured in 2PPE.  For the 3$d$ transition metals
Fe, Co, and Ni, which  show important contributions from the more
localized   $d$ states in the vicinity of the Fermi energy,
first-principles lifetime calculations in
the range of a few eV above the Fermi energy  have not, to our
knowledge, been reported in the literature so far.
  
In this paper, we present both theoretical and experimental results
for the electron dynamics as observed in 2PPE for Cu and ferromagnetic
Fe, Co, and Ni. Systematic trends among the transition metals are
discussed.  A theoretical model for the 2PPE process is presented
which is based on the time evolution of the distribution function. The
latter is calculated  with the Boltzmann equation including  effects
of photoexcitation, electron-electron scattering and
transport. Electron-electron 
scattering rates are calculated from a golden-rule expression  using
realistic DOS and constant Coulomb matrix
elements. The approach for the calculation of the scattering rates is
as outlined by Penn {\it et al.}~\cite{Penn85}  We extend this approach 
to include not only the relaxation of excited electrons, but
also the generation of secondary electrons.  Rather than performing a
first-principles calculation of the lifetime of single excited
electrons, we lay emphasis on using a model which yields
material-specific single-electron lifetimes for transition metals and
enables us to calculate the relaxation time of the distribution
including effects of secondary electrons and transport. This allows a
direct comparison of calculated relaxation times with experimental
results.

The structure of the paper is as follows. In Sec.~\ref{theory}, we
describe the model for the dynamics of excited electrons from which
2PPE is calculated. Numerical results for the relaxation of the
distribution of excited electrons are presented in
Sec.~\ref{numerical_results}.  In Sec.~\ref{experiment_section}, the
experiments are described and their results are given. In
Sec.~\ref{comparison_section}, experimental and theoretical results
are compared and discussed. Conclusions and outlook are given in
Sec.~\ref{conclusion}.

\section{Theory}
\label{theory} 
\begin{figure}[h]
\centerline{\includegraphics[width=8.6cm,angle=0]{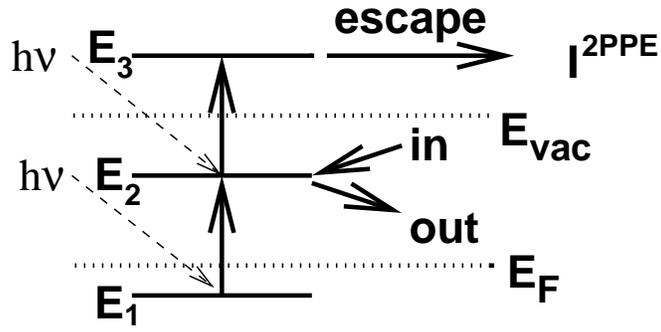}}
\vspace{1cm}
\caption{\label{2ppe_process}Illustration of the monochromatic 2PPE 
  process with
  initial state $E_1$, intermediate state $E_2$ and final state
  $E_3$. A first 
  photon excites an electron from an initial level $E_1$ in
  the range between $E_F$ and $E_F-h\nu$ into a level $E_2$.
  The population $f(E_2,t)$ depends on the
  temporal pulse
  shape of the exciting laser and is time-dependent due to
  electron-electron interaction and transport of 
  electrons out of the optically excited region into the bulk.
  A second photon excites an electron with energy $E_2$ into a
  state $E_3$ above the vacuum energy $E_{\rm vac}$, from which it can
  contribute to the 2PPE intensity via
  $I^{\rm 2PPE}(E_3,t)\propto f(E_2,t)$.} 
\vspace{1cm}
\end{figure}

The process of two-photon photoemission is illustrated in
Fig.~\ref{2ppe_process}.  The intensity $I^{\rm 2PPE}$ is obtained
by  multiplying the distribution function in the intermediate state
$f(E,\sigma,z,t)$ with a factor $e^{-z/\lambda_\sigma}$ for
transmission into the vacuum \cite{Huefner95} and with the  power of
the laser pulse $P(t)$ and integrating over time $t$ and coordinate
$z$ perpendicular to the surface:
\begin{equation}
\label{2ppe_signal}
I^{\rm 2PPE}(E+h\nu,\sigma) = \int_{-\infty}^{\infty} dt\  P(t)
  \int_0^{\infty}  dz\
e^{-z/\lambda_\sigma} f(E,\sigma,z,t) \;.
\end{equation}
Energy and spin of the intermediate state are denoted by $E$ and
$\sigma$. The photon frequency is given by $\nu$.  For the
transmission factor, we use the spin-averaged  values of the
attenuation length $\lambda$ measured in overlayer experiments for
different elements.\cite{Siegmann94}   The above expression for the
photoemission intensity based on the distribution function is suited
for the description of the population dynamics.  Our aim here is to
describe incoherent electronic processes like the decay of excited
electrons and  the generation of secondary electrons due to
electron-electron scattering. Also, the experiments with which we wish
to compare our calculations are phase-averaged measurements of the
decay of the population of excited electrons. The expression is
further justified by the fact that for bulk states in metals, one
expects rapid loss of coherence within a few fs. However, clearly, if
one is mainly interested in coherent effects like the decay of the
optically induced polarization, the treatment of the dynamics and the
photoemission process should be based on both occupation function and
polarization.\cite{Hertel96,Ogawa97,Timm99} Recent interferometric
measurements have shown relatively long decoherence times in Cu of
$T_2^\omega=5-10\rm\ fs$ for holes at the top of the $d$ bands and
electrons at about $E-E_F=1\rm\ eV$ and up to $T_2^{2\omega}=35\rm\
fs$ for electrons at about $E-E_F=4\rm\ eV$.\cite{Petek99a}

As shown in Fig.~\ref{2ppe_process}, 2PPE involves three electronic
states, in which electron-electron
scattering, electronic transport and emission into the vacuum
take place and determine the observed photoemission signal. After
optical excitation, the holes left behind in the initial state
relax and get  filled via
Coulomb scattering by electrons from occupied levels closer to the
Fermi energy. Energy conservation requires that at the same time other
electrons from below the Fermi energy are excited  to unoccupied
levels above the Fermi energy (secondary electrons). The holes are
also filled via transport processes by 
electrons from the bulk. The optically excited (primary)  electrons are
scattered out of the intermediate state by scattering with electrons
in the Fermi sea. On the other hand, secondary electrons are scattered
into the intermediate state, which leads to the refilling of this
state.  The intermediate state can be refilled by: i) an optically
excited (hot) electron after an electron-electron scattering process;
ii) a cold electron from below the Fermi energy after scattering with
a hot electron; iii) an Auger electron (an electron excited from below the
Fermi energy after a hole is filled by a cold electron). The latter
process leads to a dependence of the observed lifetime on the rate of
filling of holes (the inverse hole lifetime). The transport of excited
electrons into the bulk leads to the removal of electrons from the
intermediate state. Thirdly, the final state is above
the vacuum energy and describes a free electron which can escape from
the solid. Only electrons within a mean free path of the surface
absorbing a second photon are emitted into the vacuum.

We use now the Boltzmann equation taking into account the above
processes to describe the time evolution of the
electronic distribution function.  
The electronic states are characterized by energy $E$, spin $\sigma$,
band index $\alpha=sp,d$ and coordinate $z$ perpendicular to the
surface. The Boltzmann equation reads
\begin{equation}
\frac{\partial f(\eas z)}{\partial t}
= \left.\frac{\del f(\eas z)}{\del t}\right|^{\rm optical}
+ \left.\frac{\del f(\eas z)}{\del t}\right|_{\rm e-e}^{\rm in} 
+ \left.\frac{\del f(\eas z)}{\del t}\right|_{\rm e-e}^{\rm out}
+ \left.\frac{\del f(\eas z)}{\del t}\right|^{\rm trans}\;
\label{eom}
\end{equation}
and contains the rates of change of the occupation
due  to optical excitation, electron-electron scattering and
electronic transport. Note, the
relaxation time $\tau$ of the intermediate state which is compared
with the experimental one is determined from the decay of
the occupation. The details of the procedure are described in
App.~\ref{relaxation_time}.

For the ferromagnetic metals, we calculate the relaxation time for
excited spin-up and spin-down electrons, $\tau_\uparrow$ and
$\tau_\downarrow$. Spin-up and spin-down electrons will be denoted as
majority and minority electrons in the following.  The spin-averaged
relaxation time $\tau$ is defined as $1/\tau = 1/2\ (1/\tau_\uparrow +
1/\tau_\downarrow)$ and the relaxation time ratio as $R =
\tau_\uparrow/\tau_\downarrow$.

The optical transition rate between two electronic states due to the
interaction with the laser field with photon energy $h\nu$ is given by
\begin{equation} 
\label{optical_excitation}
\left.\frac{\del f(\eas)}{\del t}\right|^{\rm optical} = 
- \sum_{E',\beta} \left|p(E\alpha,E'\beta,\nu)\right|^2\ f(\eas)\
[1-f(E'\beta\sigma)] 
\ \rho(E'\beta\sigma)\ \delta(E-E'+h\nu) \;.
\end{equation}
Here, $p(E\alpha,E'\beta,\nu)$  is an average over electron momenta of
the optical transition matrix elements describing  the transition
between an initial occupied state in band $\alpha$ at energy $E$ and a
final unoccupied state in band $\beta$ at energy  $E'$. The DOS in the
final state is denoted by $\rho(E'\beta\sigma)$.  We use
energy-independent optical transition matrix elements.  Thus, the
strength of the optical excitation is proportional to the initial and final
DOS.  Optical excitation takes place within the optical
penetration depth of the surface and has the time dependence of the
laser field. We assume weak optical excitations,
since the energy deposited in the system is not enough to
significantly disturb the temperature or magnetization. The fraction
of excited electrons per atom is about $10^{-6}$ (see
Sec.~\ref{experiment_section}).

The transition rates due to electron-electron scattering are
calculated using Fermi's golden rule in the random-${\bf k}$ approximation,
since the strong electron-electron interaction in noble and transition
metals leads to a fast redistribution of electronic momenta so that
the information about the initial optical excitation process in ${\bf k}$
space is quickly lost. This then justifies the random-${\bf k}$
approximation for the calculation of electronic dynamics.  We extend
the treatment by Penn {\it et al.}\cite{Penn85} to a non-equilibrium
situation by calculating scattering rates into and out of a level and
taking into account the non-equilibrium distribution of electrons.
The scattering rates are derived in App.~\ref{derivation}.  The
expressions for 
the transition rates for scattering out of or into a state with energy
$E$ and spin $\sigma=\uparrow, \downarrow$ are given by:
\begin{eqnarray}
\left.\frac{\del f_{E \sigma}}{\del t}\right|_{\rm e-e}^{\rm out} &=&
- f_{E \sigma} \half  \int_{-\infty}^{\infty} dE' 
\Big\{ h_{E' \sigma} W({E \sigma},{E' \sigma})  
  + h_{E' \bar{\sigma}} W({E \sigma},{E' \bar{\sigma}}) \Big\}
  \label{out} 
\end{eqnarray}
and
\begin{eqnarray}
\left.\frac{\del f_{E \sigma}}{\del t}\right|_{\rm e-e}^{\rm in} &=&
(1-f_{E \sigma})  \half  \int_{-\infty}^{\infty} dE' 
\Big\{ e_{E' \sigma} W({E' \sigma},{E \sigma})  
 + e_{E' \bar{\sigma}} W({E' \bar{\sigma}},{E \sigma}) \Big\}\;,
 \label{in}
\end{eqnarray}
with
\begin{eqnarray}
W({E \sigma},{E' \sigma}) &=& \tph  \int_{-\infty}^{\infty} d\eps
\left( e_{\varepsilon \sigma} h_{\varepsilon+\omega,\sigma}
2 \left| M^{\uparrow\uparrow} \right|^2
+ e_{\varepsilon \bar{\sigma}} h_{\varepsilon+\omega,\bar{\sigma}}
\left|  M^{\uparrow\downarrow} \right|^2  \right)
\label{wss}
\end{eqnarray}
and
\begin{eqnarray}
W({E \sigma},{E' \bar{\sigma}}) &=& \tph  \int_{-\infty}^{\infty} d\eps
 e_{\varepsilon \bar{\sigma}} h_{\varepsilon+\omega,\sigma}
\left| M^{\uparrow\downarrow} \right|^2\;. \label{wssbar}
\end{eqnarray}
Here, $e_{E\sigma} = \rho_{E\sigma} f_{E\sigma}$ is the number of
electrons and  $h_{E\sigma} = \rho_{E\sigma} (1-f_{E\sigma})$ is the
number of holes at energy $E$, with spin $\sigma$.  The energies
involved in the transition are $E$, $E'$, $\eps$ and $\eps+\omega$,
where $\omega = E-E'$ is the energy transferred in the transition.
The spin is denoted by $\sigma$ and its opposite by
$\bar{\sigma}$. For details see App.~\ref{derivation}.

To include transport effects, we use a spatially dependent
distribution function. Due to the fact that the laser spot is much
larger than the optical penetration depth, we neglect transport in the
direction parallel to the surface and keep only the coordinate $z$ in
the direction perpendicular to the surface. The distribution function
$f(E,z)$ (suppressing spin and band indices for
simplicity) describes the occupation of an electronic state $|E\rangle$
at the coordinate $z$ . Electrons in state $|E\rangle$ are moving in 
different directions with a certain velocity distribution. In order to
describe transport, it is necessary to use
the velocity in $z$ direction as an additional argument
in the distribution function. It is now written as $f(E,z,v_z)$. 
Using Liouville's theorem, one can express the  change
in the number of electrons moving with a velocity $v_z$  as:
\cite{Ziman72}
\begin{equation}
\label{transport_eq}
\left.\frac{\del f(E,z,v_z)}{\del t}\right|^{\rm trans}
= - v_z \nabla_z\ f(E,z,v_z) \;.
\end{equation}

In the calculation of transport, we take into account the effect of
inelastic electron-electron collisions on the  distribution
$f(E,z,v_z)$. In order to do this, we describe   how both the
electron-electron scattering rates from Eqs.~(\ref{out}--\ref{in})
and the transport term from Eq.~(\ref{transport_eq}) are used in the
Boltzmann equation for $f(E,z,v_z)$.  We make the assumption that the
velocity of the electrons after scattering is randomly
distributed. The rate of change of $f(E,z,v_z)$ due to electrons
scattering into and out of this state is then given by
$\left.\frac{\del f(E,z,v_z)}{\del t}\right|^{\rm in} =
\left.\frac{\del f(E,z)}{\del t}\right|^{\rm in}$  and
$\left.\frac{\del f(E,z,v_z)}{\del t}\right|^{\rm out} =
\frac{f(E,z,v_z)}{f(E,z)} \left.\frac{\del f(E,z)}{\del t}\right|^{\rm
out},$ where the rates on the right hand side are calculated with
Eqs.~(\ref{out}--\ref{in}) using the velocity-averaged distribution
$f(E,z)=(1/N_v)\sum_{i=1}^{N_v}\ f(E,z,v_z^{(i)})$. We use $N_v$
discrete velocity intervals of width $\Delta v_z = (2 v_F)/N_v$),
neglecting the weak energy dependence of the velocity in the range of
a few eV  around $E_F$.
These terms are used in the Boltzmann equation for $f(E,z,v_z)$
together with the terms 
for transport and optical excitation. The photoexcited
electrons  have a random 
distribution of velocities: $\left.\frac{\del f(E,z,v_z)}{\del
t}\right|^{\rm optical} =  \left.\frac{\del f(E,z)}{\del
t}\right|^{\rm optical}.$

The transport effect is caused by the gradient in the particle density
created by the photoexcitation within the optical penetration
depth. We take into account the fact that $sp$ and
$d$ electrons have  different velocities due to their different degree
of localization.  The velocity of an electron in band $\alpha$ with
wave vector $k$ is given by  $v_\alpha(k) =
\frac{1}{\hbar}\frac{\partial E_\alpha(k)}{\partial k}$ Thus, nearly
free electrons in $sp$-like bands have higher velocities (or smaller
effective mass) than more localized electrons  in flat, $d$-like
bands. For $s$ electrons, we take the Fermi velocity as the maximal
velocity. For $d$ electrons, we note that the velocity in is roughly
proportional to the band width and we therefore set $v_d/v_s =
W_d/W_s$. For all elements considered, we use $v_F=18 \rm\ \AA/fs$ and
$W_d/W_s=0.4$.\cite{Moruzzi78,Harrison79} Thus, we distinguish between
different elements purely by the relative  contribution of $sp$ and
$d$ electrons. 

We have not included electron-phonon scattering into
Eq.~(\ref{eom}) because the time scale for energy transfer between
electrons and phonons is on the order of ps and thus longer than the
time scales considered in this work.\cite{Suarez95} However,
electron-phonon scattering provides an additional mechanism for
momentum transfer and can thus reduce the efficiency of ballistic
transport.\cite{Knoesel97} In our calculation, the neglect of
electron-phonon scattering might lead to an overestimate of the
transport effect. The inclusion of the momentum redistribution by
electron-phonon scattering is beyond the scope of the present work,
but will be the subject of a future publication.\cite{Knorren00} 

In Eqs.~(\ref{out}--\ref{in}), the transition rates are determined by
the available phase space for a transition weighted by the square of
the transition matrix element.   Scattering out of an excited level
and  into an excited level are treated on the same footing. The two
processes occur simultaneously because of energy conservation. The
rate  $\left.\frac{\del f(E\sigma)}{\del t}\right|_{\rm e-e}^{\rm in}$
for scattering into a level contains the effects leading to the
refilling of the intermediate state.  By calculating the scattering
rates in a consistent manner for states above and below the Fermi
energy, we keep track of the creation and relaxation of electrons as
well as holes.

Note, from the Boltzmann equation we recover the Fermi-liquid behavior
$\tau(E) \propto (E-E_F)^{-2}$ for the single-electron lifetime.
To see this we write
$\left.\frac{\del f(E\sigma)}{\del t}\right|_{\rm e-e}^{\rm out} =
-\frac{f(E\sigma)}{\tau(E\sigma)}$  which yields
$1/\tau(E\sigma) = 1/2  \int_{-\infty}^{\infty} dE'  \{ h_{E' \sigma}
W({E \sigma},{E' \sigma})   + h_{E' \bar{\sigma}} W({E \sigma},{E'
\bar{\sigma}})\}$ for the inverse lifetime.  For simplicity now we
take
$\rho=\rho_\uparrow=\rho_\downarrow$ and
$M^{\uparrow\uparrow}=M^{\uparrow\downarrow}=M$. Then the
inverse lifetime reduces to
\begin{equation}
\label{simple_lifetime}
\frac{1}{\tau(E)} = \frac{2\pi}{\hbar}\int_0^E dE' \rho(E')
\int_{-\omega}^0 d\varepsilon\ 2 \rho(\varepsilon)
\rho(\varepsilon+\omega) |M|^2\;,
\end{equation}
where $\omega = E-E'$. One sees the influence of the DOS within the
distance $(E-E_F)$ of the Fermi energy and of the Coulomb matrix
element. For energies very close to $E_F$ or for constant DOS $\rho$
one obtains
$1/\tau(E) = \frac{2\pi}{\hbar} \rho^3 |M|^2 (E-E_F)^2\;.$
Note, that the
inverse lifetime is proportional to $\rho^3$. In addition to the
factor $\rho$ from the relaxation of the initial electron at energy
$E$, one also has to take into account the available phase space for
the electron-hole pair created because of energy conservation, which
yields a factor of $\rho^2$.  This may be compared with  the
FLT lifetime expression which is
given by \cite{Quinn_Ferrell,Echenique99}
$ \tau(E) = a_0(r_s)  (E-E_F)^{-2}\;,$
where $a_0(r_s)=263 r_s^{-5/2} \rm\ fs\ eV^{2}$ and $r_s$ is the
dimensionless parameter describing the density of the electron gas. It
is given by the relation $1/n_e=4\pi (r_s a_0)^3/3$, where $n_e$ is
the electron density and $a_0$ is the Bohr radius. Thus, the
expression for $\tau(E)$ derived from  $\left.\frac{\del
f(E\sigma)}{\del t}\right|_{\rm e-e}^{\rm out}$ in the appropriate
limit gives the same energy dependence as FLT.

One can try to understand the relaxation times observed in different metals
by means of the FLT expression for $\tau$ by using
$r_s$ corresponding to the electron density in the metal. However, the
expression is strictly valid only for  free-electron-like metals and
it is not simple to extend it to take into account $d$
electrons. First one can use the electron density corresponding to
$sp$ electrons only. We take values for integrated $sp$ and $d$
electron densities of states in the solid from
Ref.\onlinecite{Papa86}. The number of $sp$ electrons per atom for Fe,
Co, Ni, and Cu is  1.07, 1.13. 1.03 and 1.10, respectively.  One
obtains roughly $r_s \sim 2.6$ and  $a_0 \sim 25\rm\ fs\ eV^{2}$ for
all the metals Fe, Co, Ni, and Cu. This neglects $d$ electrons
altogether and does not yield any differences between these metals. On
the other hand, if one uses the total  number of $sp$ and $d$
electrons, one obtains values for $r_s$ monotonically decreasing from
from $r_s=1.32$ in Fe to  $r_s=1.22$ in Cu, leading to  $a_0=130\rm\
fs\ eV^{2}$  for Fe and $a_0=161\rm\ fs\ eV^{2}$ for Cu.  It is
evident from the magnitude of $a_0$ that this overestimates the
influence of  $d$ electrons. Clearly a more refined treatment is
necessary which distinguishes between $sp$ and $d$ electrons and takes
into account the DOS within a few eV of the Fermi energy. The
influence of the $d$ bands on the lifetime of a state depends on the
distance $E_d$ of the $d$ bands to   the Fermi energy: for small
excitation energy $\Delta E=E-E_F< E_d$, the $d$ bands should have
little influence on the lifetime. In Sec.~\ref{comparison_section}, we
will discuss the influence of $d$ bands in more detail.

\section{Numerical Results of the Theory}
\label{numerical_results}
To show clearly the
influence of different physical mechanisms,  the calculated relaxation
times are presented in three steps  including consecutively more
processes in the calculation.

In the first step, we consider the single-electron lifetime. At this
level, we take into account in the Boltzmann equation,
Eq.~(\ref{eom}), only scattering  out of a particular level in
addition to the optical excitation. Results
are labelled by ({\it out}) in 
Figs.~\ref{cu36comp} and \ref{standard_comparison}. After
photoexcitation, the time evolution of the distribution function shows
an exponential decay. The lifetime obtained in this way is a
single-electron lifetime and can be compared with FLT  or with
lifetimes obtained from  first-principles calculations of the
self-energy.\cite{Campillo99,Schoene99} However, the single-electron
lifetime is observed in an experiment only if there are no other
effects present such as secondary-electron generation or
transport. Thus, the calculated single-electron lifetime should not be
directly compared with  experimental results.  It serves as a guide
for comparison with other theoretical results and as a reference for
comparison with results when secondary electrons and transport are
included.

In the second step, we take into account secondary electrons,
while  neglecting transport effects. Thus, in the Boltzmann equation,
we keep the scattering terms for scattering {\it
into} and {\it out of} the levels. Results are labelled  ({\it out,
in}). The relaxation time obtained in this way  includes effects of
the whole distribution. Note, it is not a 
single-electron lifetime, but an effective relaxation time of the
distribution.

In the third step, we also take into account transport effects. The
loss of excited 
electrons due to transport out of the surface region will influence
the occupation and hence the apparent electronic relaxation time.
Results are labelled ({\it out, in, transport}).

The DOS for the different metals used as input in the calculation of
the scattering rates in Eqs.~(\ref{out}--\ref{wssbar})   are taken
from an FLAPW calculation \cite{Dewitz98} and are shown in
Fig.~\ref{dos}. We distinguish only between $d$-like and $sp$-like
states in the  DOS. We take the partial $d$ DOS from the calculation
and use the remaining DOS as $sp$-like DOS. 
The total DOS are very similar to the ones given in
Ref.\onlinecite{Moruzzi78}.  For Cu, however, we shift the $d$ bands
to lower binding energy by 0.4 eV in order to  obtain agreement with
ARPES results for the binding energy of the $d$ bands.\cite{Huefner95}
It is known that LAPW calculations might yield too small binding
energies for the $d$ bands.\cite{Eckhardt84}

The values for the optical penetration depth were
obtained from the optical constants in Ref.\onlinecite{Johnson72}. We
use $\lambda_{\rm Fe}=124$~\AA, $\lambda_{\rm Co}=108$~\AA,
$\lambda_{\rm Ni}=122$~\AA, and $\lambda_{\rm Cu}=149$~\AA\ for
$h\nu=3.0$~eV. For the range of photon energies considered, the energy
dependence of the penetration depth can be neglected.

\subsection{Numerical Results for Cu} 
\label{cu_section}
\begin{figure}[h]
\begin{center}
\includegraphics[width=8.6cm,angle=0]{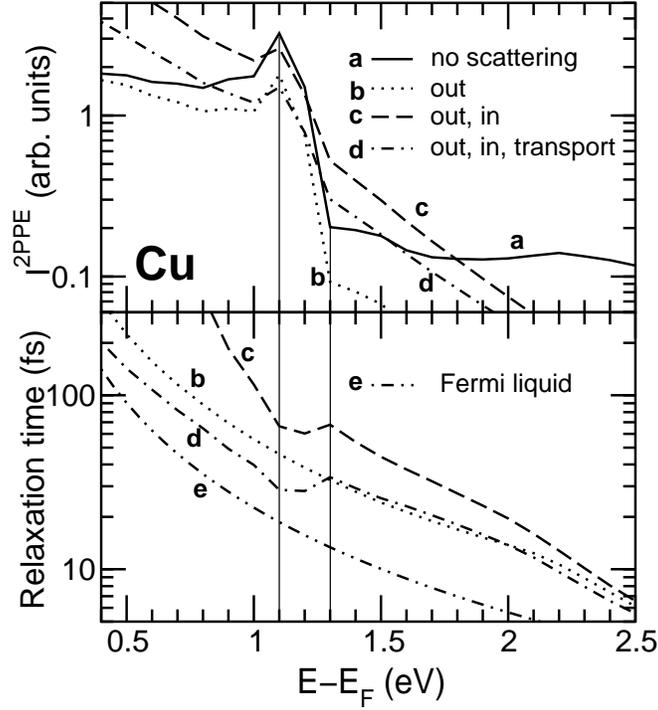}
\end{center}
\vspace{1cm}
\caption{\label{cu36comp}Calculated 2PPE intensity 
  and relaxation time of the excited electron distribution for Cu for
  photon energy $h\nu = 3.3 \rm\ eV$.   Curve {\it a} shows the 2PPE
  intensity if no scattering is present and reflects the distribution
  of optically excited electrons. Curve {\it b} gives the result if
  only scattering out of the intermediate level is kept in
  Eq.~(\ref{eom}). Curve {\it c} is the result if scattering into the
  intermediate state (secondary electron effect) is also included.
  Curve {\it d} represents  the case when the effect of transport is
  also taken into account.  The relaxation time when only scattering
  out of the intermediate state is kept (curve {\it b}) is a
  single-electron lifetime and can be compared with the lifetime
  predicted by Fermi-liquid theory, shown in curve {\it e}. The other
  relaxation times are effective relaxation times of the distribution
  of excited electrons.}
\vspace{1cm}
\end{figure}

Fig.~\ref{cu36comp} illustrates the effect of secondary electrons and
transport on the intensity $I^{\rm 2PPE}$ (upper part) and relaxation
time (lower part) for optical excitation  with photon energy
$h\nu=3.3\rm\ eV$. The results for Cu show particularly clearly the
influence of secondary electrons and transport. We use  $M=0.8\rm\ eV$
and $\left|M^{\uparrow\uparrow}/M^{\uparrow\downarrow}\right|=1$ for
the Coulomb matrix elements. This choice will be justified  at the end
of this section.

In curve {\it a} of Fig.~\ref{cu36comp}, we show $I^{\rm 2PPE}$
calculated from Eq.~(\ref{2ppe_signal}) without scattering and
transport. In this case $I^{\rm 2PPE}$ is proportional to the number
of optically excited (primary) electrons in the intermediate
state. The optical excitation in Eq.~(\ref{optical_excitation}) is
proportional to the convolution of initial and final DOS, since we use
constant optical transition matrix elements. The intensity shows an
important contribution from initial states in the $d$ band below
$E-E_F=1.3\rm\ eV$ with a pronounced peak at $1.1\rm\ eV$ and a small
contribution from initial states in the $s$ band above  $1.3\rm\ eV$.

In curve {\it b} of Fig.~\ref{cu36comp}, we show $I^{\rm 2PPE}$
keeping only the out-scattering term in Eq.~(\ref{eom}), which
corresponds to an exponential decay of the optically excited
distribution.  The intensity is reduced compared to the intensity
without scattering, and the reduction gets stronger towards higher
excitation energy due to the shorter lifetime. The relaxation time
calculated in this way is a single-electron lifetime. For energies
below $E-E_F=2\rm\ eV$, it shows the energy dependence   $ \tau(E)
\propto (E-E_F)^{-2}$ as in FLT. The FLT lifetime for Cu with
$a_0=25\rm\ fs\ eV^{2}$ is shown by curve {\it e}. It is a factor of
about 2.5 lower than the lifetime calculated using $M=0.8\rm\ eV$.

In curve {\it c} of Fig.~\ref{cu36comp}, we show  the results using
the in-scattering term as well as the out-scattering term in
Eq.~(\ref{eom}). At low energy a secondary-electron tail forms and the
intensity becomes a superposition of the  initial optical excitation
and the  secondary-electron tail.  The relaxation time shown by curve
{\it c} is an effective relaxation time of the distribution including
secondary-electron effects. It is no longer monotonous and  shows a
distinct feature in the region of the intensity peak.  We find a
relative minimum at the position of the intensity peak. Further, the
relative maximum  corresponds to the $d$-band threshold above which
very few optically excited $d$ electrons are found. We can understand
this by studying the contribution of secondary electrons. Comparing
the intensities without (curve {\it b}) and with (curve {\it c})
secondary electrons, we find that in the region of the peak, the
intensity is only slightly increased by secondary electrons, whereas
it is strongly increased  above  the $d$-band threshold. This shows
that in the intensity peak one observes mainly relaxation of optically
excited electrons, whereas above the threshold, one finds mainly
contributions of secondary electrons. Comparing the relaxation times,
we find that in the region of the intensity peak and also at high
energies $E-E_F>2.5\rm\ eV$, the relaxation time with secondary
electrons (curve {\it c}) comes very close to the relaxation time
without secondary electrons (curve {\it b}), again showing that one
observes  mainly relaxation in these regions. Most secondary electrons
are generated in the process of filling the holes in the $d$ band
created by the optical excitation (so-called Auger effect).  The
increase in the relaxation time is due to the fact that secondary
electrons are generated with a certain delay after the optical
excitation corresponding to the $d$-hole
lifetime.\cite{Knoesel98,Knorren99}

In Fig.~\ref{cu36comp} curve {\it d} shows the results obtained using
transport in addition to secondary-electron effects. Compared to the
case without transport (curve {\it c}), the intensity  is reduced. The
transport effect removes excited particles from the observation region
close to the surface into the bulk and thus reduces the intensity. The
relaxation time including transport (curve {\it d}) has a similar
shape as the relaxation time with secondary-electron effects (curve
{\it c}), but the magnitude of the relaxation  time is strongly
reduced by transport. Interestingly,  curve {\it d} comes close to
curve {\it b} in the region above the threshold ($E-E_F>1.3 \rm\ eV$).
Thus, we find that for Cu in a certain energy range, the effects of
secondary electrons and transport on the relaxation time roughly
cancel. This  has been pointed out in an analysis of a 2PPE experiment
before.\cite{Knoesel97}

Fig.~\ref{cu_all_comp} shows theoretical and experimental results for
the 2PPE intensity and relaxation time for photon energies $h\nu=3.0$
and $3.3 \rm\ eV$.  Both the intensity peak and the relaxation time
feature shift linearly with photon energy. As indicated
by the dotted lines,  the minimum in the relaxation time corresponds
to the peak in the intensity, whereas the maximum in the relaxation
time corresponds to the $d$-band threshold.

We have used a Coulomb matrix element $M=0.8\rm\ eV$. With this choice
we obtain at $E-E_F=1\rm\ eV$ for the single-electron lifetime
$\tau=55\rm\ fs$ and for the relaxation time including secondary
electron and transport effects $\tau=40\rm\ fs$ (for photon energy
$h\nu=3.3\rm\ eV$).  The single-electron lifetime is in good agreement
with first-principles calculations  in the energy range $E-E_F<2\rm\
eV$.\cite{Campillo99}  It also yields good agreement with
experimentally determined relaxation times at low energies
($E-E_F\approx 1\rm\ eV$).\cite{Knoesel98}  We remark that at higher
energies ($E-E_F>2\rm\  eV$), we would obtain better agreement with
first-principles calculations and experiments for a smaller Coulomb
matrix element $M=0.6\rm\ eV$. Thus, the use of energy-dependent
matrix elements might improve the overall agreement between theory and
experiment over a wide energy range.

\subsection{Numerical Results for Fe, Co, and Ni}
\label{transition_section}
\begin{figure}[h]
\begin{center}
\vspace{1cm}
\includegraphics[width=8.6cm,angle=0]{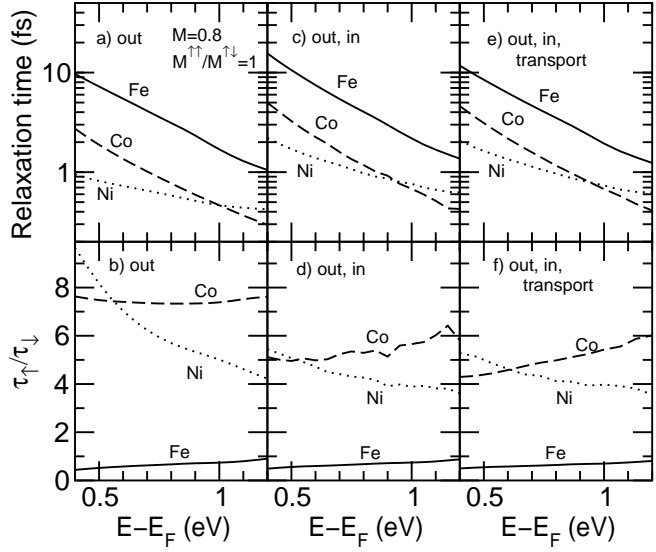}
\vspace{1cm}
\end{center}
\caption{\label{standard_comparison} Spin-averaged
  relaxation time of the distribution and ratio of relaxation times
 for majority and minority electrons.  Results labelled {\it out}  are
 single-electron lifetimes. Results ({\it out,in}) refer to relaxation
 times of the distribution including secondary electrons.  Results
 ({\it out, in, transport}) also include transport.  The same average
 Coulomb matrix element $M=0.8\rm\ eV$ and
 $|M^{\uparrow\uparrow}/M^{\uparrow\downarrow}|=1$ is used.}
\vspace{1cm}
\end{figure}

First, we discuss numerical results shown in
Figs.~\ref{standard_comparison}a and b and labelled by ({\it out}) for
the single-electron lifetime in  Fe, Co, and Ni. We use
the same energy-independent Coulomb matrix element $M=0.8\rm\ eV$ and
$\left|M^{\uparrow\uparrow}/M^{\uparrow\downarrow}\right|=1$ for the
different metals.

\begin{figure}
\begin{center}
\includegraphics[width=8.6cm,angle=0]{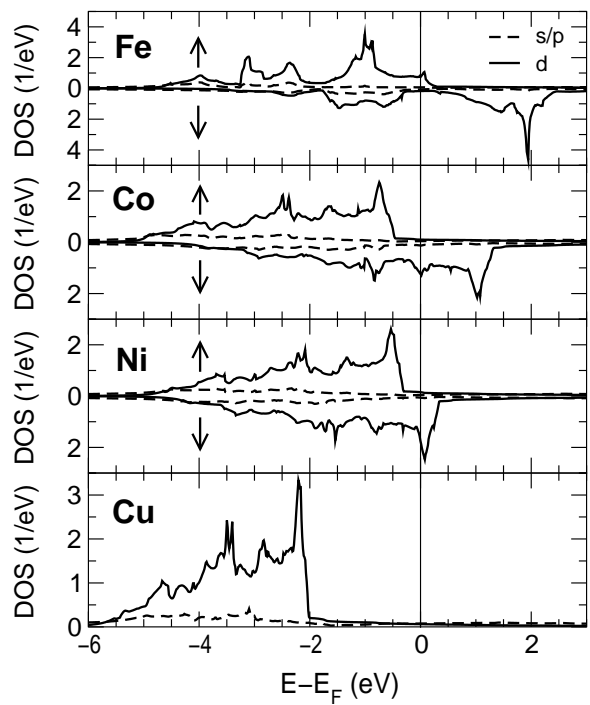}
\end{center}
\vspace{1cm}
\caption{\label{dos}LAPW density of states (DOS) used as input in
  the calculation of the electron-electron scattering rates in
  Eqs.~(\ref{out}--\ref{wssbar}). }
\vspace{1cm}
\end{figure}

For constant and equal Coulomb matrix element $M$, the single-electron
lifetime is directly related to the DOS shown in Fig.~\ref{dos}
and used as input for the calculation.  The influence of the DOS on
the lifetime  is seen in the scattering-rate
expression Eq.~(\ref{out}) or in the simplified expression
Eq.~(\ref{simple_lifetime}).  The scattering rate, the inverse of the
lifetime, is proportional to a combination of terms which contain
products of three factors of the DOS. 

When comparing the single-electron lifetime of Cu shown in
Fig.~\ref{cu36comp} (curve  {\it b}) with the results in
Fig.~\ref{standard_comparison}a for the transition metals,  one can
see that Cu has a lifetime much longer than the other metals. This is
due to  the small total DOS close to $E_F$.  In Cu, the $d$ bands are
located about 2~eV below  $E_F$ and there is only a very small total
DOS close to  $E_F$ (Fig.~\ref{dos}).  The small $d$ DOS close to
$E_F$ is due to hybridization with $sp$-like  states.  In Ni with one
electron less than Cu, the $d$ bands move closer to $E_F$.
Furthermore, the $d$ bands are split into a minority and majority spin
band and  a small portion of the minority $d$ band is unoccupied,
extending up to about 0.4~eV above  $E_F$ (Fig.~\ref{dos}).  Due to
the pronounced peak at the upper edge of the $d$-band, the minority
DOS close to $E_F$ is extremely large.  This leads to a large phase
space for electron-electron scattering at low energy. Thus at low
energy (below $E-E_F=1\rm\ eV$), Ni has the smallest calculated
single-electron lifetime among the four elements.   Co, with one
electron less than Ni, has an even larger portion of unoccupied
minority DOS, extending up to about 1.2~eV above  $E_F$. Although the
total number of unoccupied states is higher in Co  than in Ni, the
minority DOS at $E_F$ is lower in Co. Thus at low energy (below 1~eV),
Co has less phase space  and the calculated lifetime is longer  than
in Ni. With increasing energy, more and more unoccupied states become
available in Co, so that above 1~eV, the calculated lifetime in Co
becomes shorter than the one in Ni.  In Fe, again with one electron
less compared to Co, the unoccupied minority DOS extends up to 2.4~eV
above $E_F$, and even the majority DOS has a small unoccupied
fraction. The minority DOS at $E_F$ in Fe is lower than in Co and in
Ni, so that Fe has the  smallest phase space and the longest
calculated lifetime at low energy.

Therefore, at low energy (below $E-E_F=1\rm\ eV$), our simplified
theory with equal $M$ for the different metals gives   $\tau_{\rm
Ni}<\tau_{\rm Co}<\tau_{\rm Fe}$. This trend  changes for Co and Ni
above 1~eV. Then we get $\tau_{\rm Co}<\tau_{\rm Ni}$. At even higher
energy (above 2 eV, not shown in the figure), all of the unoccupied
$d$ states in Fe are available for a transition and one calculates $
\tau_{\rm Fe}<\tau_{\rm Co}<\tau_{\rm Ni}$.  This relation is also
observed in transmission experiments above the vacuum energy for
electrons with energies above $E-E_F=5\rm\ eV$.\cite{Siegmann92}

In Fig.~\ref{standard_comparison}b  the ratio of majority to minority
single-electron lifetime $\tau_\uparrow/\tau_\downarrow$ is shown.
For Co, this ratio  is nearly constant with a value
$\tau_\uparrow/\tau_\downarrow=7.5$.  This is  understandable in view
of the high and nearly constant ratio of  minority to majority DOS at
low energy. For Ni the ratio decreases from
$\tau_\uparrow/\tau_\downarrow=9.5$ at $E-E_F=0.4\rm\ eV$ to
$\tau_\uparrow/\tau_\downarrow=4$ at $E-E_F=1.4\rm\ eV$. The  decrease
is due to  the fact that above 0.4~eV there are no more unoccupied
minority $d$ states. The additional phase space gained by going to
higher energy is the same for minority and majority electrons, leading
to a smaller ratio. In Fe  the ratio increases from
$\tau_\uparrow/\tau_\downarrow=0.5$ to 1 for excitation energies
between $E-E_F=0.4$ and $1.2\rm\ eV$.  Thus, majority electrons have a
shorter calculated lifetime than minority electrons at low
energy. This results from the unoccupied portion of the majority DOS
above $E_F$, which for low energy leads to a larger phase space for
the relaxation of majority electrons and therefore to a shorter
lifetime.

Secondly, we discuss results shown in Figs.~\ref{standard_comparison}c
and d obtained when secondary electrons are included in the
calculation. They are labelled by ({\it out, in}).  Transport effects
are still neglected.    The inclusion of
secondary electrons leads to an increase of the relaxation time by a
factor of about two as compared to the single-electron lifetime.  The
strongest effect is found at the lowest energies. The increase is
stronger for the elements with the shortest  calculated lifetimes (Ni,
Co), so that the differences in the relaxation time including
secondary electrons between Ni, Co, and Fe are smaller than for the
single-electron lifetime. The ratio $\tau_\uparrow/\tau_\downarrow$
for Ni and Co is reduced to $\tau_\uparrow/\tau_\downarrow=4-5$ for Ni
and $\tau_\uparrow/\tau_\downarrow=5-6$ for Co. For Fe, the ratio is
nearly unchanged,  $\tau_\uparrow/\tau_\downarrow=0.5-1$.  The
reduction of the ratio $\tau_\uparrow/\tau_\downarrow$  due to
secondary electrons is understandable in view of the fact that the
inclusion of secondary electrons leads to a coupling between majority
and minority electron populations via electron-electron
scattering. Relaxing minority electrons can excite majority electrons
and vice versa. For Co and Ni, for example, longer-living majority
electrons will continue to excite minority electrons after the
shorter-living primary minority electrons have relaxed. The apparent
minority-electron relaxation time becomes longer  and the ratio
$\tau_\uparrow/\tau_\downarrow$ becomes smaller by this process.  The
trends among the calculated relaxation times of the  transition
metals, particularly the relation $\tau_{\rm Ni}<\tau_{\rm
Co}<\tau_{\rm Fe}$ at $E-E_F<1\rm\ eV$ are unchanged when secondary
electrons are included.

The spectral shape of the optical excitation,  i.e.~the distribution
of primary electrons, has some influence on  the calculated relaxation
time.  For example, electrons excited to a high energy lead to more
secondary electrons (due to the short lifetime of the primary
electrons) and to a distribution extending to higher energy (due to
the high energy of the primary electrons). Similar arguments apply to
the energetic position of the holes created by the optical excitation.
The results for the transition metals shown in
Fig.~\ref{standard_comparison} are  obtained using an optical
excitation with photon energy $h\nu = 3.0\rm\ eV$ and constant optical
transition matrix elements in Eq.~(\ref{optical_excitation}). We have
also studied the effect of a different  optical excitation on the
relaxation time. We have modelled  a resonant optical excitation where
excitations take place dominantly  between an initial state at the top
of the $d$ band and a final state in the $sp$ band. No significant
difference in the calculated relaxation time for the two different
shapes of the excitation is obtained.

Thirdly, we discuss  results of the calculations including transport
in addition to secondary-electron effects in the Boltzmann equation,
Eq.~(\ref{eom}). They are labelled by ({\it out, in,
transport}). Results are  shown   in Figs.~\ref{standard_comparison}e
and f.   The inclusion of transport leads only to a very slight
reduction of the effective relaxation time in Fe, Co, and Ni.  The
ratio $\tau_\uparrow/\tau_\downarrow$ as well is  only slightly
affected by transport.  This is in contrast to  the strong reduction
of the effective relaxation time in Cu (compare curves {\it c} and
{\it d} in Fig.~\ref{cu36comp}).  The explanation is that
the typical transport relaxation timescale is about
40~fs.\cite{Aeschlimann96,Knoesel98}  This is shorter than, or
comparable 
to, the single-electron lifetime in Cu in the energy range considered,
but considerably longer than the typical lifetimes in Fe, Co, or Ni.
Thus, transport is expected to have great influence for Cu, but
not for the transition metals.

\section{Experiment}
\label{experiment_section}
\subsection{TR-2PPE technique}
The TR-2PPE pump-probe experiments are carried out in a UHV chamber by
monitoring the number of electrons at a given kinetic energy as a
function of the delay between the pump and probe pulses. We employ the
equal-pulse correlation technique, i.e.~the two pulses are
monochromatic and equal in intensity, but cross-polarized. For metals,
the use of orthogonal linear polarized light pulses suppresses
coherent interference effects (within the limit of rapid dephasing) to
a large extent.\cite{Aeschlimann96}  Otherwise they influence the
optical transition process and would make the reconvolution of the raw
data much more difficult. Furthermore, the influence of Cs-induced
surface states on the lifetime can be suppressed (see below).

The non-linearity of the two-photon process leads to an increase in
the 2PPE yield when the pulses are spatially and temporarily
superimposed. As long as the two laser pulses temporarily overlap it
is obvious that an electron can be emitted by absorbing just one
photon from each pulse. However, if the pulses are temporarily
separated, then an excited electron from the first pulse is able to
absorb a photon from the second pulse but only as long as the
inelastic lifetime of the intermediate state exceeds the delay or the
normally unoccupied electronic state is refilled by a secondary
electron. Due to a precise measurement of the time delay between the
two pulses (1~fs $\hat{=}$ path length difference of 0.3~$\rm\mu m$),
this technique  allows us to analyze relaxation times which are
considerably shorter than the laser pulse duration.

We use laser pulses at low fluence and peak power to avoid
space-charge effects or highly excited electron distributions. We
emphasize  that the count rate is much lower than one electron per
pulse. Therefore, we measure the relaxation and transport of
individual excited electronic states rather than the collective
behavior of transiently heated non-equilibrium distribution. We have
to roughly calculate the fraction of excited electrons. Typically, we
have a laser fluence of about 0.3~nJ/pulse in each beam resulting in
$6\times10^8$ photons per pulse. For a spot size of $\sim 150 \rm\ \mu
m$ and a penetration  depth of the blue light of $\sim 150\rm\ \AA$,
the volume in which the laser light  will be absorbed is about
$3\times10^{-10}\rm\ cm^3$. If 10\% of the light is absorbed  by the
metal, then $6\times10^7$ photons are absorbed by $7\times10^{13}$
atoms which  results in a fractional excitation of roughly 1 part in
$10^6$.

\subsection{Experimental set-up}
\begin{figure}[h]
\begin{center}
\includegraphics[width=8.6cm,angle=0]{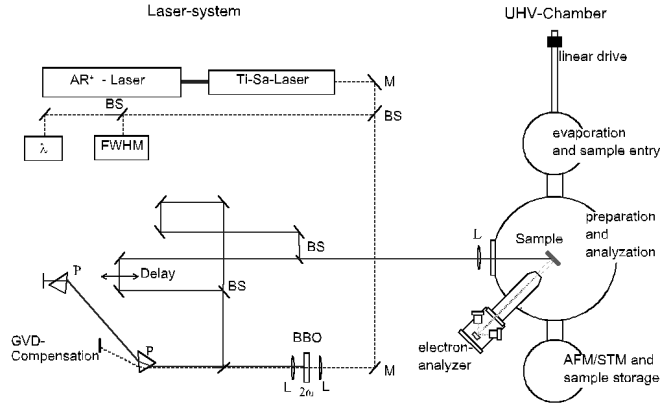}
\end{center}
\vspace{1cm}
\caption{\label{setup}Schematic view of the equal pulse correlation
  set-up for time resolved 2PPE.}
\vspace{1cm}
\end{figure}

A schematic overview of the experimental set-up is shown in
Fig.~\ref{setup}. The samples are mounted in a UHV chamber with a base
pressure in the $10^{-11}\rm\ mbar$ range. It is equipped with a
cylindrical sector electron energy analyzer (CSA) and a spin analyzer,
based on spin-polarized low-energy electron diffraction
(SPLEED).\cite{b}  The earth's magnetic field is shielded by
$\mu$-metal coverings  inside the chamber. Standard surface-physics
methods such as Auger-electron spectroscopy (AES) and low-energy
electron diffraction (LEED)  are available to check the cleanliness
and the surface structure of the samples. The orientation of the
samples is $45^\circ$ with respect to the laser beam and the electrons
are detected in normal-emission geometry. Remanent magnetization of
the ferromagnetic samples is achieved by a magnetic field pulse from a
coil. The geometric arrangement of the spin analyzer allows the
measurement of the spin polarization along the horizontal in-plane
axis of the sample. To minimize the effects of stray fields and to
facilitate electron collection, a bias voltage ($-4\rm\ V$ for Cu and
$-15\rm\ V$ for Ni, Co, and Fe)  is applied between the sample and the
CSA. For spin-resolved measurements the electrons are guided into the
SPLEED analyzer, which is located on top of the CSA. In this spin
analyzer the electrons are first accelerated to 104.5 eV kinetic
energy, as the highest figure of merit for this kind of analyzer is
known to be at this primary-electron energy.\cite{b}  Thereafter, the
electrons are scattered at a  tungsten (001) crystal. From the
resulting LEED feature, the (-2,0) and (2,0) electron beams that have
a high spin asymmetry, are counted in two different channeltrons. The
Sherman factor S, a quantity for the spin selectivity of an analyzer,
is found to increase from 0.2 to 0.25 over the two years of operation;
the highest value was reached after having the tungsten SPLEED crystal
a long period in very low pressure.

The time-resolved 2PPE experiments are performed with a femtosecond
mode-locked Ti:sapphire laser, pumped by about 10 W from a cw Ar$^+$
laser. The system delivers transform-limited and sech$^2$ temporal
shaped pulses of up to 9 nJ/pulse with a duration of 40 fs at a
repetition rate of 82 MHz. The linearly polarized output of the
Ti:Sapphire laser is frequency-doubled in a 0.2 mm thick Beta Barium
Borate (BBO) crystal to produce UV pulses at $h\nu$ = 3 to 3.4 eV. The
UV beam is sent through a pair of prisms to pre-compensate for pulse
broadening due to dispersive elements like lenses, beamsplitters and
the UHV-chamber window in the optical path. A GVD and intensity-loss
matched interferometric autocorrelator set-up is used for the
pump-probe experiment (see Fig.~\ref{setup}). The pulses are split by
a beamsplitter to equal intensity (pump and probe pulses), and one
path is delayed with respect to the other by a computer-controlled
delay stage. Both beams are combined co-linearly but cross-polarized
by a second beamsplitter and are focused at the sample surface.

For the ferromagnetic samples, we use evaporated films because they
can be held magnetized in a single-domain state without an applied
external field and the stray field is much smaller compared to a bulk
ferromagnet. In principle, this experiment could also be performed
with bulk samples. The ferromagnetic films are evaporated onto a
Cu(001) substrate in a separable chamber. We use a water-cooled
evaporator based on electron-beam heating. The material to be
evaporated (of 99.999\% purity) was inside a molybdenum crucible (Co,
Fe) or directly evaporated from a 1 mm thick wire (Ni). The
evaporation rate (around 0.2 nm/min) was checked with a quartz
oscillator, which is calibrated against atomic-force-microscope
thickness measurements. During evaporation the pressure remained in
the $10^{-10}\rm\ mbar$ region.

The thickness of the ferromagnetic films must fulfill the following
requirements. Firstly, it has to be large enough, in order to avoid an
influence of electrons from the Cu substrate. Secondly, it should be
possible to remanently magnetize the film by a suitable strong field
pulse. And thirdly, the axis of the magnetization has to lie in the
film plane because the geometry of our spin analyzer only allows the
measurement of transversally polarized electrons.

Cobalt films, epitaxially grown on Cu(001) surfaces, are formed in a
stable fcc structure and exhibit in-plane magnetization. For thick
films, the in-plane magnetization easy axis lies along the (110)
direction of the Cu crystal.\cite{c}  In-plane magnetization was
detectable starting at a thickness of around 0.4 nm. This is in
agreement with other investigations using the magneto-optical Kerr
effect.\cite{d}  Above a film thickness of 2 nm, the spin polarization
did not increase any more on further deposition of Co. For our
investigations we evaporated 10 nm thick Co films.

Iron grows in the fcc structure on Cu(001) during the initial steps of
evaporation. The magnetization vector is oriented perpendicular to the
film surface. Above a thickness of around 2 nm a bcc (110) structure
starts developing.\cite{e}  The magnetic easy axis for these thick
films is found to lie parallel to the Cu(100) axis.\cite{f}  We use 20
nm thick iron films for our investigation.

For Ni on Cu(001) the magnetization vector is in-plane for small
thicknesses, then it switches to out-of-plane at a thickness of around
1.2 nm.\cite{g}  Only at larger thicknesses of around 6-7 nm does it
turn back to in-plane.\cite{h}  We found a saturation of the spin
polarization for thicknesses  above 25 nm. Therefore, we evaporated 40
nm thick Ni films for our measurements.

The clean metal surfaces are first dosed with Cs to lower the surface
work function, a well-known technique. This enabled lifetime
measurements of lower excited states, increasing the useful energy
range of the spectra closer to the Fermi energy (see
Fig.~\ref{2ppe_process}). Cs is evaporated from a thoroughly
out-gassed getter source  (SAES). The effect on the lifetime by dosing
a metal surface with small amount of Cs ($<$0.1 ML) has been
thoroughly investigated in the last years.\cite{i,j}  Using
cross-polarized pulses no differences in the lifetime have ever been
found between a clean and a cesiated metal surface by means of
TR-2PPE.\cite{k}  In addition, we found no differences in spin
polarization between the clean and the cesiated surfaces in the
overlapping energy region between 1.7 and 3.3~eV.

\subsection{Experimental results for Cu}
\label{exp_cu_results}
In Fig.~\ref{cu_all_comp}b the extracted relaxation time as a function
of the intermediate-state energy for a Cu(111) surface is shown, using
a photon energy of 3.0~eV ($\Box$) and 3.3~eV ({\LARGE $\circ$}). The
data are  reconvoluted  from the experimentally obtained
cross-correlation traces using a rate-equation model for the
population of the intermediate state. In the  case of rapid dephasing,
and assuming an exponential depletion of the nascent photoexcited
electron population, the evolution of the transient population
$N^*(t)$ of the intermediate state is given by $dN^*(t) /dt = A(t) -
N^*(t)/\tau$ where $A(t)$ is the excitation induced by the first
(pump) laser pulse. Details of the way to extract the relaxation time
from the measured signal have been given in previous
publications.\cite{Schmuttenmaer94,Aeschlimann96,Knoesel98}  As
expected, the lifetime increases as the excited state energy
decreases, caused by the reduced phase space for electron-electron
scattering (see Secs.~\ref{theory}, \ref{cu_section}, and Fermi-liquid
theory). However, at an  intermediate-state energy, where the
intensity becomes dominated by interband transition from the $d$ band
to the unoccupied $sp$ band (strong $d$-band peak in the intensity),
the measured relaxation time decreases by more than a factor of two
before it increases again. By changing the photon energy from $h\nu =
3.0\rm\ eV$ to $h\nu = 3.3\rm\ eV$, both the $d$-band peak in the
intensity and the dip in the measured lifetime move with the same
energy difference $\Delta E = \Delta h\nu$ as expected. The striking
difference in the values observed in the energy range $E - E_F =
0.7-1.3\rm\ eV$ indicate quite clearly that the relaxation time can
depend critically on the used photon energy.

\subsection{Experimental results for Fe, Co, and Ni}
\subsubsection{Spin-integrated time-resolved 2PPE measurements}
\begin{figure}
\begin{center}
\includegraphics[width=8.6cm,angle=0]{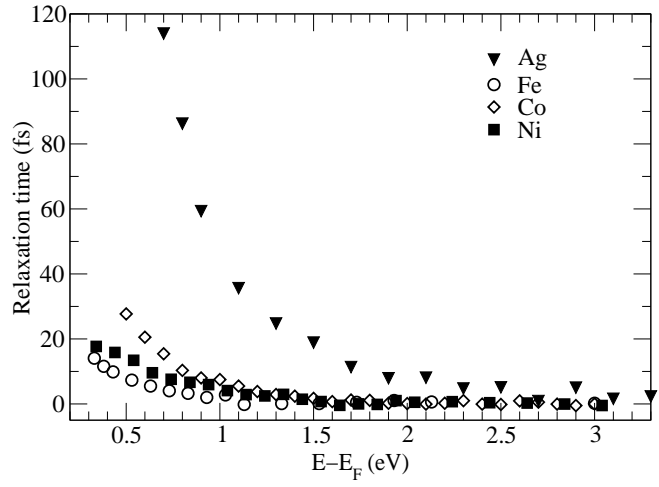}
\end{center}
\vspace{1cm}
\caption{\label{aeschli_ave}Comparison of experimental relaxation time
  results for Ag and the three transition metals, Fe, Co, and Ni.}
\vspace{1cm}
\end{figure}

We used the same equal-pulse correlation technique to extend the
investigation of the hot-electron relaxation to transition metals Co,
Fe, and Ni. Compared with noble metals, in which the $d$ shell is
completely filled (see Cu and Ag), the $d$ band of  transition metals
is only partially filled, and the electronic and relaxation properties
are dominated to a considerable degree by these $d$ electrons. The
strong localization of these $d$ electrons results in a narrower band
and hence in a much higher DOS near the Fermi level as compared with
the $sp$ electrons in Cu and Ag. A higher density  of occupied and
unoccupied states near the Fermi level is expected to lead to faster
relaxation and hence to a shorter inelastic lifetime of excited
electronic states as discussed in Sec.~\ref{transition_section}. This
prediction is well satisfied by our data. Fig.~\ref{aeschli_ave} shows
a comparison of  the extracted relaxation time of silver and the three
investigated ferromagnetic transition metals cobalt, nickel and
iron. The experimental values for these metals are at least a factor
of 10 smaller than those of Cu and Ag. These small values reduce the
energy range in the region of 0.3~eV to 1.3~eV which provides a
meaningful statement about the relation in the relaxation time between
the three investigated transition metals. In contrast to the numerical
results calculated using the same Coulomb matrix element $M$ for
different metals (see Sec.~\ref{transition_section}), the data
indicate a relation $\tau_{\rm Fe} < \tau_{\rm Ni} < \tau_{\rm Co}$
within this energy range.

\subsubsection{Spin- and time-resolved 2PPE measurements}
Adding a spin analyzer to the CSA energy analyzer makes possible the
separate but simultaneous measurement of both spin states. The
electrons of a fixed energy are counted according to their spin in two
different channeltrons as a function of the time delay between the two
pulses, at a given magnetization direction. To compensate for an
apparatus-induced asymmetry, the magnetization is then reversed and
the measurement is taken again. From the resulting four datasets the
relaxation times $\tau_\uparrow$ and $\tau_\downarrow$ for spin-up and
spin-down electrons are extracted by using the same reconvolution
method as discussed above. Each pair of data points presented in the
plots of this section is the average of eight to ten single relaxation
time measurements.

The spin dependence will have a superimposed effect on the
spin-integrated relaxation time. Therefore, a spin dependence in the
relaxation time can only be resolved if there is already a certain
relaxation time detectable with spin-integrated measurements. As shown
in Fig.~\ref{aeschli_ave}, in the energy range above 1.4~eV, we find
for all three  transition metals a relaxation time smaller than our
time resolution ($<$2~fs). On the other hand, at intermediate-state
energies close to $E_F$, the electrons emitted by 1PPE processes start
becoming important. They induce a large background to the 2PPE signal
and make an accurate extraction of the lifetimes difficult. Therefore,
spin-resolved measurements can only be usefully performed for
intermediate-state energies between 0.3 and 1.1~eV.

\begin{figure}
\begin{center}
\includegraphics[width=8.6cm,angle=0]{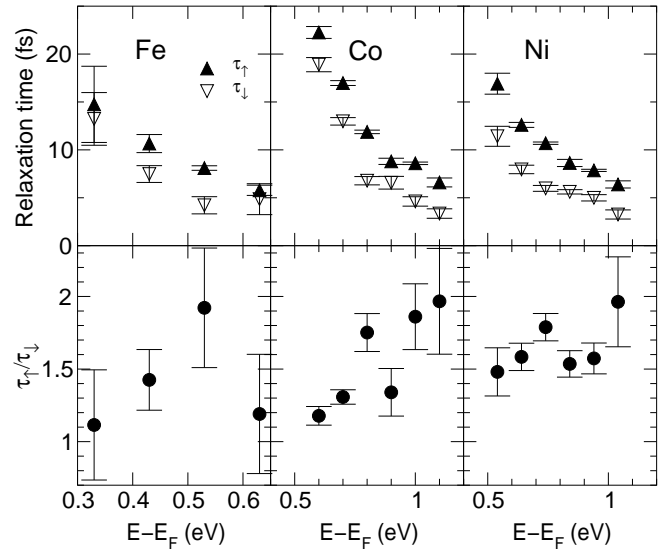}
\end{center}
\vspace{1cm}
\caption{\label{aeschli_spin}Experimental results for the
  spin-resolved relaxation time $\tau_\uparrow$ and $\tau_\downarrow$
  and the ratio $\tau_\uparrow/\tau_\downarrow$ 
  of majority to
  minority relaxation time.}
\vspace{1cm}
\end{figure}

In Fig.~\ref{aeschli_spin} the spin-dependent relaxation time for
electrons (upper  part) and the ratio of majority to minority lifetime
(lower part) of Fe, Co, and Ni films are plotted. The error bars in
the plot represent the statistical scatter. The experimental results
of the three examined ferromagnetic materials show two common facts:
i) The lifetime for majority-spin electrons is always found to be
longer than the lifetime for minority-spin electrons and ii) the value
for $\tau_\uparrow/\tau_\downarrow$ was  found to lie between 1 and
2. The largest differences between $\tau_\uparrow$ and
$\tau_\downarrow$ are  found for Ni and Co, whereas for Fe, the
difference is slightly reduced. This qualitative behavior of the
spin-dependent lifetime can be readily explained by the excess of
unfilled minority-spin states compared to unfilled majority-spin
states. According to this simple model, the spin dependence of the
scattering rate is larger for the strong ferromagnets Co and Ni than
for the weak ferromagnet Fe. This is in agreement with our
measurements, where only a small spin effect could be detected for
Fe. In Fe, this model would even predict a reversal of the effect for
low energies below 1~eV, i.e.~the lifetime for spin-down electrons
should become longer than the lifetime for spin-up electrons, see
Fig.~\ref{standard_comparison}b. A ratio of majority to minority
relaxation time $R$  below 1 is, however, not observed for $E - E_F <
1\rm\ eV$. This result indicates that the simple model, considering
the different number of empty electronic states as the only decisive
factor for a spin-dependent relaxation time, is not sufficient for a
quantitative interpretation of our experimental data.

\section{Discussion}
\label{comparison_section}
\begin{figure}[h]
\begin{center}
\includegraphics[width=8.6cm,angle=0]{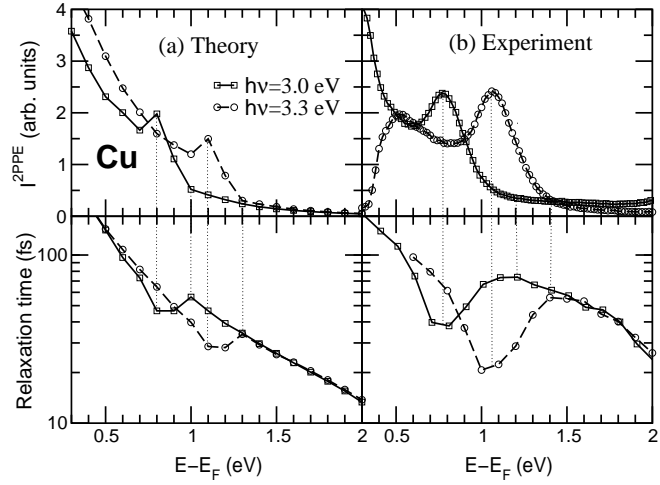}
\end{center}
\vspace{1cm}
\caption{\label{cu_all_comp}Calculated (a) and measured
  (b) 2PPE intensity and relaxation time for Cu for photon energies
  $h\nu = 3.0$ and $3.3 \rm\ eV$. Calculations include secondary
  electrons and transport.  Note the dependence of the relaxation time
  on photon energy, especially in the region of the peak in the
  intensity ($E-E_F = 0.7 - 1.3 \rm\ eV$). The minimum in the
  relaxation time corresponds to the peak in the intensity and the
  maximum in the relaxation time corresponds to the $d$-band
  threshold, as indicated by the dotted lines, in agreement with
  several experiments.\cite{Pawlik97,Cao97,Knoesel98}}
\vspace{1cm}
\end{figure}

First, in Fig.~\ref{cu_all_comp} we compare  experimental and
theoretical results for Cu for photon energies $h\nu=3.0$ and $3.3\rm\
eV$.  Experimental and theoretical results show qualitative agreement
regarding the main features. The peak in the intensity and the feature
in the  relaxation time (relative minimum and maximum) shift linearly
with photon energy. The minimum in the relaxation time corresponds to
the peak in the intensity and the maximum in the relaxation time
corresponds to the $d$-band threshold. The explanation that the
feature in the calculated relaxation time is due to the secondary
electrons  was given in detail in Sec.~\ref{cu_section}. The good
agreement between calculated and experimental results is a strong
evidence for this explanation. The calculations reproduce the features
observed in different experiments \cite{Pawlik97,Cao97,Knoesel98} in a
natural way by including secondary electrons without invoking further
explanations such as excitonic states involving 3$d$
electrons.\cite{Cao97}  The differences between theoretical and
experimental results in Fig.~\ref{cu_all_comp} lead to several
conclusions.  Both the peak in the intensity and the  difference
between the minimum and maximum of the relaxation time  are more
pronounced in the experiment than in the calculation. This may be an
evidence that the calculation yields too many secondary electrons
which cause a too strong background  and thus a too small structure in
the 2PPE intensity (see the upper part of
Fig.~\ref{cu_all_comp}a). The relatively small increase from the
minimum to the maximum in the calculated relaxation time should not be
affected much by this, but may rather point to  the fact that the
lifetime of the $d$ holes in the calculation is too small. Note, due
to the large $d$ DOS below $E_F$, there is no symmetry between hole
lifetimes and lifetimes of excited electrons. In our calculations,
hole lifetimes for energies below the $d$-band threshold are very
short due to the large DOS. If the hole lifetimes were larger, then
fewer secondary electrons would be generated, but with a longer delay
after the creation of the electron-hole pair by the laser pulse. A
longer delay will lead to the observation of a longer relaxation time
in the excited state when secondary-electron contributions are
important.

The interpretation given here  that the non-monotonous feature in the
lifetime in Cu is due to secondary (Auger) electrons has raised some
controversy in the
literature.\cite{Pawlik97,Cao97,Knoesel98,Petek97,Petek99} While
Knoesel {\it et al.}\cite{Knoesel98} interpret their data by
contributions from  Auger electrons at intermediate-state energies
above the $d$-band peak, Petek {\it et al.}\cite{Petek99} argue that
secondary electrons make no significant contribution to the signal
above $E-E_F=1.5\rm\ eV$ (for photon energy  $h\nu=3.1\rm\ eV$). The
argument is based on a temperature-dependent delayed rise in the 2PPE
signal as a function of time delay between the laser pulses, which is
observed below and immediately above the $d$-band peak at
$E-E_F=0.9\rm\ eV$, but which  vanishes above $E-E_F=1.5\rm\ eV$ and
in the region of the peak.  The temperature-dependent delayed rise is
interpreted as a contribution from Auger electrons, which  agrees with
our interpretation of the feature in the lifetime.  In
Ref.\onlinecite{Petek99}, the fact that the delayed rise vanishes
above 1.5~eV is taken as evidence that Auger electrons are absent in
this energy region. In contrast to this conclusion, our calculations
show significant contributions from secondary  electrons up to about
$E-E_F=2.5\rm\ eV$ (compare the relaxation times without [curve $b$]
and with [curve $c$] secondary electrons in  Fig.~\ref{cu36comp}). We
argue in the following that the absence of a resolvable second peak in
the 2PPE signal  is no evidence for the absence of Auger
electrons. Thus the results of Ref.\onlinecite{Petek99} are not in
contradiction with our results.  One would observe a second peak with
a delay given by the hole lifetime at the $d$-band peak if all Auger
electrons were created at a fixed rate corresponding to this hole
lifetime.  However, Auger electrons are also created by the filling of
holes deeper in the $d$ band with energies up to $h\nu$. These deep
holes have have shorter lifetimes than the ones at the top of the $d$
band. Thus they lead to secondary-electron contributions to the
dynamics in the intermediate state with  a smaller delay time. The
fact that holes with different lifetimes contribute to the
secondary-electron dynamics makes it difficult to observe a resolvable
second peak with a fixed delay corresponding to the lifetime at the
$d$-band peak.  Thus in our view, the measurements reported in
Refs.\onlinecite{Petek97,Petek99} are not in contradiction with the
interpretation of the non-monotonous feature in the relaxation time
given by us and in Ref.\onlinecite{Knoesel98}.

\begin{figure}[h]
\begin{center}
\includegraphics[width=8.6cm,angle=0]{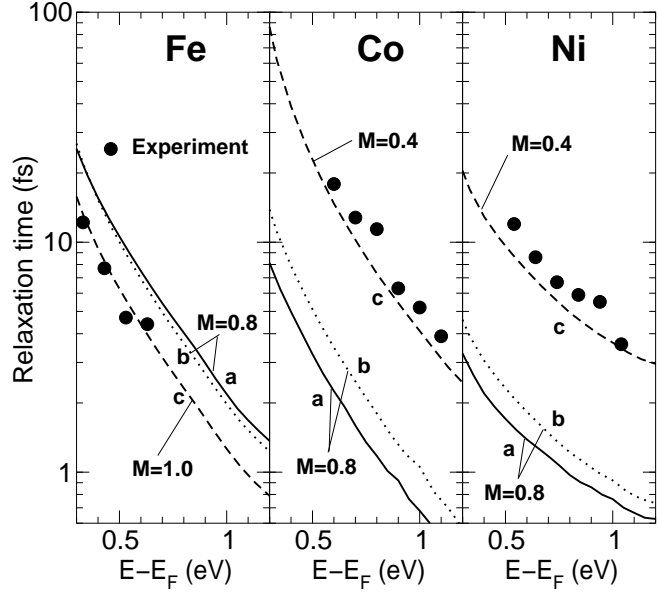}
\end{center}
\caption{\label{tau_comparison} Experimental and theoretical results
  for the spin-averaged relaxation time of the
  distribution. Calculations include secondary electron effects. Curve
  {\it a} shows results using Coulomb matrix element $M=0.8\rm\ eV$
  and $|M^{\uparrow\uparrow}/M^{\uparrow\downarrow}|=1$ for the
  various transition metals. Curve {\it b} gives results for the same
  $M$, but
  $|M^{\uparrow\uparrow}/M^{\uparrow\downarrow}|=0.5$. Results using
  different values of $M$ for the various  transition metals and
  $|M^{\uparrow\uparrow}/M^{\uparrow\downarrow}|=0.5$  are shown in
  curve {\it c}.}
\end{figure}

\begin{figure}[h]
\begin{center}
\includegraphics[width=8.6cm,angle=0]{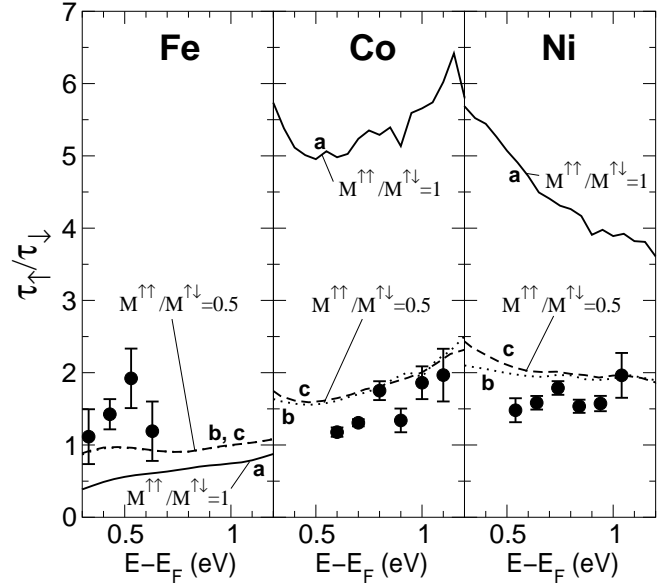}
\end{center}
\caption{\label{ratio_comparison}
  Experimental and theoretical results for the
  ratio $\tau_\uparrow/\tau_\downarrow$ of the relaxation 
  time of the distribution. The labels are as in
  Fig.~\ref{tau_comparison} and refer to the same parameters.}
\end{figure}

In Figs.~\ref{tau_comparison} and \ref{ratio_comparison} we compare
experimental and theoretical results for the spin-averaged relaxation
time $\tau$ and the ratio $\tau_\uparrow/\tau_\downarrow$ of majority
and  minority relaxation time for the ferromagnetic transition metals
Fe, Co, and Ni.  The discrepancies between  experimental and
theoretical results indicate that both DOS and Coulomb matrix elements
play a role.  Note, theoretical  results refer to relaxation times of
the distribution including secondary-electron effects. Transport
effects have been neglected here in view of the fact that they cause
only minor changes in the relaxation time  of the transition metals,
see Fig.~\ref{standard_comparison}.

First, the results calculated with $M=0.8\rm\ eV$ for all the
transition metals are shown by the curves  {\it a} in
Figs.~\ref{tau_comparison} and \ref{ratio_comparison}.  The difference
in the calculated relaxation times  for Fe, Co, Ni, and Cu is then
only due to the different DOS used as input for the calculation. Note,
the calculated relaxation time is smaller than the experimental one in
Co and Ni, while it is larger in Fe. The calculated ratio
$\tau_\uparrow/\tau_\downarrow$ is larger in Co and Ni than the
experimenal one, but it is smaller in Fe.

Secondly, in curves {\it b}, we show results of calculations using
again $M=0.8\rm\ eV$, but the reduced value
$\left|M^{\uparrow\uparrow}/M^{\uparrow\downarrow}\right|=0.5$. One
expects that the matrix element $M^{\uparrow\uparrow}$ for scattering
of parallel spins is smaller than $M^{\uparrow\downarrow}$ for
antiparallel spins, since electrons with parallel spins avoid each
other due to the Pauli exclusion principle.\cite{Friedel69}  The ratio
$\tau_\uparrow/\tau_\downarrow$ is strongly reduced in Co and Ni,
while it is increased in Fe, which leads to satisfactory agreement for
$\tau_\uparrow/\tau_\downarrow$  between experimental and theoretical
results. The spin-averaged relaxation time is not strongly affected by
the value of
$\left|M^{\uparrow\uparrow}/M^{\uparrow\downarrow}\right|$.

Thirdly, we take into account different Coulomb matrix elements $M$
for the various metals, while we still use
$\left|M^{\uparrow\uparrow}/M^{\uparrow\downarrow}\right|=0.5$.  The
results are given by the curves {\it c} in Figs.~\ref{tau_comparison}
and \ref{ratio_comparison}.  For Co and Ni we use $M=0.4\rm\ eV$,
while for Fe we take $M=1.0\rm\ eV$.  The use of these values for $M$
leads to reasonable agreement between theoretical and experimental
results for both the spin-averaged relaxation time and the ratio
$\tau_\uparrow/\tau_\downarrow$.

Different Coulomb matrix elements in Fe, Co, Ni, and Cu are mainly
caused by the influence of $d$ electrons.  Note, while in isolated
atoms, Coulomb matrix elements do not vary much from Cu to
Fe,\cite{Mann67} in solids the band character, the position of the $d$
band, and the screening of $d$ electrons are expected to change
this. The screened Coulomb matrix elements for scattering between
Bloch states with wave vectors ${\bf k}_i$ in bands $\alpha_i$ is
given by
\begin{equation}
M^{1;2}_{3;4}=\int\ d^3 r\ d^3 r'\  \psi^*_{{\bf k}_1 \alpha_1}({\bf
r}) \psi^*_{{\bf k}_2 \alpha_2}({\bf r}')
\frac{e^2}{\varepsilon(|{\bf r}-{\bf r}'|,\omega)\ |{\bf r}-{\bf r}'|}
\psi_{{\bf k}_3 \alpha_3}({\bf r})  \psi_{{\bf k}_4 \alpha_4}({\bf
r}') \;.
\end{equation}
Here, $\varepsilon$ is the dielectric function. The Coulomb matrix
elements are of course influenced by $d$ electrons, since their wave
functions are more localized and also since they contribute to the
screening of the Coulomb  potential.  The strong localization of $d$
electrons leads to smaller overlap with $sp$-electron wave functions
and therefore  to smaller transition matrix elements when $sp \to d$
transitions are involved as compared to matrix elements involving $sp
\to sp$ transitions. Note, the $d$-electron wave functions get more
localized from Fe to Cu.  The additional screening of $d$ electrons is
contained in the dielectric function $\varepsilon(|{\bf r}-{\bf
r}'|,\omega)$,  where $\omega$ is the energy transfered in the
transition. Somewhat depending on $\omega$, $d$ electrons closer to
the Fermi energy contribute mainly to screening. In the static limit
($\omega\to 0$), the Lindhard dielectric function  for a free electron
gas reduces to $\varepsilon(q)=1+k_0^2/q^2$ such that the screened
Coulomb interaction in real space takes the form
$V(r)=\frac{e^2}{r}e^{-k_0 r}$. In the  Thomas-Fermi approximation,
the screening wave vector is directly related to the DOS at the Fermi
level,\cite{Ziman72} $k_0=4\pi e^2\rho(E_F)$.  In the case of
transition metals, the expressions are not strictly valid because  $d$
electrons are not free-electron-like.  Although the quantitative
contribution of $d$ electrons to screening is not well-known,
qualitatively it is clear that a higher DOS near the Fermi level leads
to stronger screening. This may explain that the screened Coulomb
matrix element in Co and Ni with many $d$ electrons close to the Fermi
energy is smaller than the one in Cu with nearly no $d$ electrons
close  to the Fermi energy. For larger energy transfer $\omega$, also
electrons further away from the Fermi energy contribute to
screening. Then ultimately the total number of $d$ electrons
influences screening. This could be the reason why Fe, which has fewer
$d$ electrons, has a larger Coulomb matrix element than Co, Ni, and Cu.

After completion of our study we became aware of related work about
the influence of $d$ electrons on the lifetime of low-energy electrons
in noble and transition
metals.\cite{Zarate99,Campillo99,Schoene99} Choosing different matrix
elements for  $sp$ and $d$ states, Zarate {\it et al.}\cite{Zarate99}
obtain good agreement with experimental results for
Co.\cite{Aeschlimann97} Campillo {\it et al.}\cite{Campillo99} and
Sch{\"o}ne  {\it et al.}\cite{Schoene99} have calculated lifetimes in
Cu using the density-functional theory for the determination of the
electronic structure and have found important contributions of $d$
electrons to the lifetime via screening, localization of the wave
function and DOS effects.

\section{Conclusion}
\label{conclusion}
We have presented experimental and theoretical results for the
dynamics of excited electrons in Cu, Fe, Co, and Ni.

The results for Cu show the influence of secondary electrons and
transport effects on the  observed relaxation time. The non-monotonous
behavior in the relaxation time obtained in the calculation is in
qualitative agreement with experiments. It seems desirable to achieve
a better quantitative agreement for this structure in order to draw
definitive conclusions about the role played by secondary electrons.

Experimental results for the spin-dependent relaxation time for  Fe,
Co, and Ni reveal that $\tau_{\rm Fe}<\tau_{\rm Ni}<\tau_{\rm Co}$ and 
that $\tau_\uparrow/\tau_\downarrow$ lies between 1 and 2
for the three metals.

The comparison of experimental and theoretical results shows that  DOS
effects alone do not explain the magnitude of the relaxation time for
the various transition metals observed in experiments. The differences
between the calculation using the same Coulomb matrix element for Fe,
Co, and Ni and experiments  reveal that Coulomb-matrix-element effects
are important.

As an outlook for further studies, we conclude that more detailed
calculations have to include a first-principles calculation  of the
Coulomb interaction matrix elements. Notably one has to take into
account the screening by the $d$ electrons, the influence  of the
localized $d$-electron wave functions on the Coulomb matrix elements,
and the energy dependence of the matrix elements.  Further studies of
the influence of the optical excitation, for example the influence of
the photon energy and the polarization of the incoming light on the
observed dynamics, are needed. Also the influence of the hole lifetime
on the calculated relaxation times  should be investigated.  The rate
of filling of $d$-band holes influences the time evolution of the
distribution and hence also the observed relaxation time for levels
above the Fermi energy via the generation of secondary electrons.
Hole lifetimes have been observed in recent experiments on
Cu,\cite{Matzdorf99} and  their effect on two-photon photoemission has
also been considered in a recent theoretical work.\cite{Sakaue99}

\acknowledgments R.K.\ and K.B.\ acknowledge  financial
support by Deutsche Forschungsgemeinschaft (SFB 290). We wish to
thank J.~Dewitz for providing results of band-structure calculations
and C.~Timm and G.~Bouzerar for interesting discussions. R.B.\ and M.A.\
would like to thank all co-workers who contributed to the experimental
work described here, in particular M.~Bauer, S.~Pawlik, W.~Weber, and
D.~Oberli. H.C.~Siegmann is thanked for many stimulating discussions.

\appendix
\section{Determination of the Relaxation Time} 
\label{relaxation_time}
\begin{figure}[h]
\begin{center}
\includegraphics[width=8.6cm,angle=0]{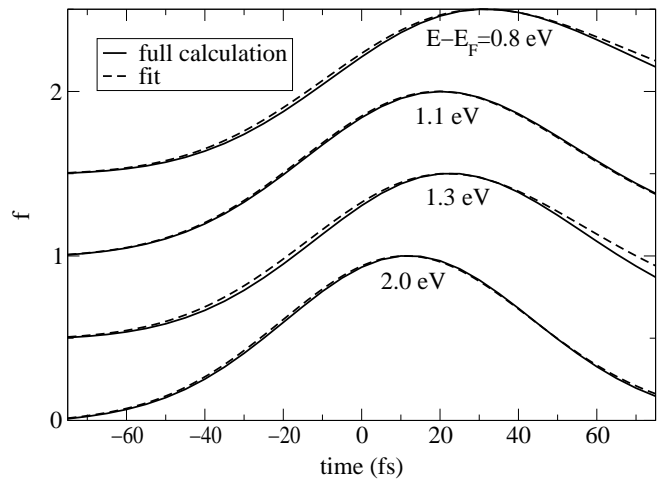}
\end{center}
\vspace*{0.5cm}
\caption{\label{fit}
  Calculated occupation function for different intermediate state
  energies (normalized to 1 at the maximum) and fit by a function
  describing exponential decay. The 
  calculation is for Cu with $h\nu = 3.3$~eV and a pump laser of 70~fs
  duration.}
\end{figure}

In order to determine the effective relaxation time, we fit the 
occupation function calculated from Eq.~(\ref{eom})
with a function $f$ describing exponential decay with a
relaxation time $\tau$ obtained from the equation: $ \frac{\partial
f}{\partial t} = \left.\frac{\del f}{\del t}\right|^{\rm
optical} - \frac{f}{\tau}\;.$ The fit to the occupation
calculated from Eq.~(\ref{eom}) is done by taking
$\tau$ for which $f$ has the maximum at the same time. The
comparison of the curves in  Fig.~\ref{fit} shows that this is a valid
procedure over a wide range of energies, even where secondary
electrons dominate (see $E-E_F=0.8$ or 1.3~eV). The fact that the
deviations 
are very small shows that  although the calculated curves do not
exactly show exponential behavior, they can be fitted well by a curve
$f$ showing exponential decay for some effective relaxation time.

\section{Electron-electron Scattering Rates} 
\label{derivation}
The scattering rate out of the state with momentum ${\bf p}$, band
$\alpha$ (designating $sp$ or $d$-like wave function) and spin
$\sigma$ is given in first order time-dependent perturbation theory
(golden rule) by:
\begin{eqnarray}
\left.\frac{\del f_\pas}{\del t} \right|_{\rm e-e}^{\rm out} &=&
  -f_\pas \tph \sum_{{\bf p}'{\bf kk}',  \beta\gamma\delta}  \bigg\{
  \half f_\kgs (1-f_\ppbs)(1-f_\kpds)  \left| \mps - \mks \right|^2
  \nonumber\\  && \times \dels \nonumber\\ && +  f_\kgsb
  (1-f_\ppbs)(1-f_\kpdsb) \left| \mpsb \right|^2 \nonumber\\ && \times
  \delsbp  \bigg\} \;.
\end{eqnarray}
In the same manner, one defines the scattering rate 
$\left.\frac{\del f_\pas}{\del t} \right|_{\rm e-e}^{\rm in}$
for scattering into a state.
The first and second terms describe scattering between electrons of
of the same and of opposite spin, respectively.

Sums over momenta are converted into integrals over
energies in the random-{\bf k}
approximation.\cite{Penn85} 
The conversion to {\bf k}-averaged
quantities is done in the following way:
\[
\sum_{\bf k} f_\kas \rightarrow \int_{-\infty}^{\infty} dE \rho_\eas
f_\eas \;.
\]
Each  {\bf k} sum leads to a factor of the  DOS, $\rho_\eas$. Products
of distribution functions and densities of 
states are defined as $e_\eas = \rho_\eas f_\eas$ and $ h_\eas =
\rho_\eas (1-f_\eas)$
and designate the number of
electrons and holes.
The random-{\bf k} approximation allows
the replacements ${\bf p}\to E$, ${\bf p}'\to E'$, ${\bf k}\to \eps$
and ${\bf k}'\to \eps'$. 
The integral over $\eps'$ can be performed because of the
$\delta$-function and allows to replace $\eps'$ by $\eps + \omega$
with $\omega=E-E'$.
The {\bf k}-averaged expressions for the scattering rates
out of and into the state  $\eas$ are then given by:
\begin{eqnarray}
\left.\frac{\del f_\eas}{\del t}\right|_{\rm e-e}^{\rm out} &=& -
f_\eas \half \sum_{\beta} \int_{-\infty}^{\infty}  dE'  \Big\{ h_\epbs
W(\eas,\epbs)   + h_\epbsb W(\eas,\epbsb) \Big\}\;,
\label{out_rate_appendix}\\ \left.\frac{\del f_\eas}{\del
t}\right|_{\rm e-e}^{\rm in} &=& (1-f_\eas)  \half \sum_{\beta}
\int_{-\infty}^{\infty}  dE'  \Big\{ e_\epbs W(\epbs,\eas)   +
e_\epbsb W(\epbsb,\eas) \Big\}\;. \label{in_rate_appendix}
\end{eqnarray}
with the definitions
\begin{eqnarray}
W(\eas,\epbs) &=& \tph \sum_{\gamma\delta} \int_{-\infty}^{\infty}
d\eps \left( e_\epsgs h_\epspds \left| \mes - \mepss \right|^2 \right.
\nonumber\\   &&\left.  + e_\epsgsb h_\epspdsb \left| \mesb \right|^2
\right) \;, \label{w_noflip_appendix} \\  W(\eas,\epbsb) &=& \tph
\sum_{\gamma\delta} \int_{-\infty}^{\infty}  d\eps e_\epsgsb h_\epspds
\left| \mepssb \right|^2\;. \label{w_flip_appendix}
\end{eqnarray}

For the calculation, we will use a simple parametrization
of the matrix elements.
First, we do not further distinguish between
states of $sp$ or $d$ symmetry in the matrix element.  
Then the form of
Eqs.~(\ref{out_rate_appendix}-\ref{w_flip_appendix}) 
remains the same when the partial DOS is replaced by the
total DOS.
Secondly, we neglect the energy dependence, 
assuming it to be weak
in the range of a few eV from the Fermi energy. 
In Eq.~(\ref{w_noflip_appendix}), we neglect the interference term
after expanding the modulus square: 
\begin{equation}
\left| M_1^{\sigma\sigma} - M_2^{\sigma\sigma} \right|^2 = \left|
M_1^{\sigma\sigma} \right|^2  + \left| M_2^{\sigma\sigma} \right|^2 -
[M_1^{\sigma\sigma} (M_2^{\sigma\sigma})^* + \rm{c.c.}]  \approx
\left| M_1^{\sigma\sigma} \right|^2 +  \left| M_2^{\sigma\sigma}
\right|^2 = 2 \left| M^{\sigma\sigma} \right|^2\;.
\end{equation}
We have denoted the matrix elements with different energy arguments by
$M_1^{\sigma\sigma}$ and $M_2^{\sigma\sigma}$. In the last step, we
drop the energy index and use energy-independent matrix elements
$M^{\sigma\sigma} = M_1^{\sigma\sigma} = M_2^{\sigma\sigma}$.  The
approximation of neglecting the interference term was also made by
Penn {\it et al.}\cite{Penn85}  The Coulomb matrix elements for
scattering between equal and opposite spins are denoted by
$M^{\sigma\sigma}$ and $M^{\sigma\bar{\sigma}}$. We use
$M^{\uparrow\uparrow}=M^{\downarrow\downarrow}$ and
$M^{\uparrow\downarrow}=M^{\downarrow\uparrow}$.  For the calculation,
we define two parameters for the average matrix element squared and
for the ratio of $M^{\uparrow\uparrow}$ and $M^{\uparrow\downarrow}$:
$ M^2 = \frac{\left|M^{\uparrow\uparrow}\right|^2 +
\left|M^{\uparrow\downarrow}\right|^2}{2}\;, m  =
\frac{\left|M^{\uparrow\uparrow}\right|}
{\left|M^{\uparrow\downarrow}\right|} \;.$

The simplified equations used in the calculations of the
scattering rates are given in Eqs.~(\ref{out}--\ref{wssbar}) .

\clearpage

\newpage 

\end{document}